\documentclass[sigconf,screen=true,bookmarks=false,9pt]{acmart}
\newcommand\blfootnote[1]{%
  \begingroup
  \renewcommand\thefootnote{}\footnote{#1}%
  \endgroup
}
\usepackage{graphicx}

\usepackage{amsmath}
\usepackage{mathtools}
\usepackage{comment}
\usepackage[subrefformat=parens,labelformat=parens]{subfig}
\usepackage{bm}
\usepackage{multirow}
\usepackage{diagbox}
\usepackage{multicol}
\usepackage{titlesec}
\usepackage{threeparttable,booktabs}
\usepackage{tikz}
\usepackage{balance}
\usepackage{courier}                                       % courier font, used in \texttt
\usepackage{cleveref}                                      % smart citation
\usepackage[mathcal]{eucal}
\usepackage[noend]{algpseudocode}
\algrenewcommand\textproc{\texttt}
\makeatletter\let\float@addtolists\relax\makeatother
\usepackage{algorithm}
\usepackage{filecontents}                                  % support to pgfplots
\usepackage{pgfplots}
\usepackage{pgfplotstable}
\pgfplotsset{compat=newest}
\usepackage{siunitx}
\theoremstyle{plain}

\theoremstyle{definition}

\usepackage[skip=1pt]{caption}            % set space between figure and caption
\graphicspath{{./figs/}}
\begin{document}
\settopmatter{printacmref=false} % Removes citation information below abstract
\renewcommand\footnotetextcopyrightpermission[1]{} % removes footnote with conference information in first column

\title{
\textbf{MORE--Stress: \underline{M}odel \underline{O}rder \underline{R}eduction based \underline{E}fficient Numerical Algorithm for Thermal \underline{Stress} Simulation of TSV Arrays in 2.5D/3D IC} \vspace{-10pt}}
\author{ 
    Tianxiang Zhu$^{1}$,
    Qipan Wang$^{1,2}$,
    Yibo Lin$^{1,3,4*}$, 
    Runsheng Wang$^{1,3,4}$, 
    Ru Huang$^{1,3,4}$ \\ \Large
    \mbox{$^1$School of Integrated Circuits, 
   $^2$Academy for Advanced Interdisciplinary Studies, Peking University, Beijing,} \\
    $^3$Institute of Electronic Design Automation, Peking University, Wuxi,\\
   $^4$Beijing Advanced Innovation Center for Integrated Circuits\\
    { \{txzhu,\ qpwang,\ yibolin,\ r.wang,\ ruhuang\}@pku.edu.cn}
\vspace{-5pt}}
\begin{abstract}
    Thermomechanical stress induced by through-silicon vias (TSVs) plays an important role in the performance and reliability analysis of 2.5D/3D ICs. While the finite element method (FEM) adopted by commercial software can provide accurate simulation results, it is very time- and memory-consuming for large-scale analysis. Over the past decade, the linear superposition method has been utilized to perform fast thermal stress estimations of TSV arrays, but it suffers from a lack of accuracy. In this paper, we propose MORE-Stress, a novel strict numerical algorithm for efficient thermal stress simulation of TSV arrays based on model order reduction. Extensive experimental results demonstrate that our algorithm can realize a 153--504$\times$ reduction in computational time and a 39--115$\times$ reduction in memory usage compared with the commercial software ANSYS, with negligible errors less than 1$\%$. Our algorithm is as efficient as the linear superposition method, with an order of magnitude smaller errors and fast convergence. 
\end{abstract}

\maketitle
\blfootnote{$^*$Corresponding author}

\titlespacing*{\section}{0pt}{1.2ex plus .0ex minus .0ex}{.3ex plus .0ex}
\vspace{-25pt}
\section{Introduction}
\label{sec:intro}
2.5D/3D ICs are one of the most promising technologies to achieve the increasingly demanding integration and performance targets \cite{chang2024physical}. Nevertheless, many reliability issues in 2.5D/3D ICs remain challenging, among which the thermomechanical stress induced by the mismatch in thermal expansion coefficients and the thermal load between the room temperature and fabrication temperature plays an important role. Severe thermal stress will incur mechanical cracking and damage, accelerate degradation, and affect the mobility of transistors, leading to degraded performance \cite{moroz20243dic,jung2012chip,chen2023combined,liang2020electromigration}. Generally, the thermal stress in 2.5D/3D ICs is evaluated by finite element method (FEM) in commercial software like ANSYS \cite{ansys} for verification and design optimization \cite{pathak2011electromigration,pak2012electromigration,lau2018warpage}.
\begin{figure}[t]
    \centering
    \includegraphics[width=0.3\textwidth]{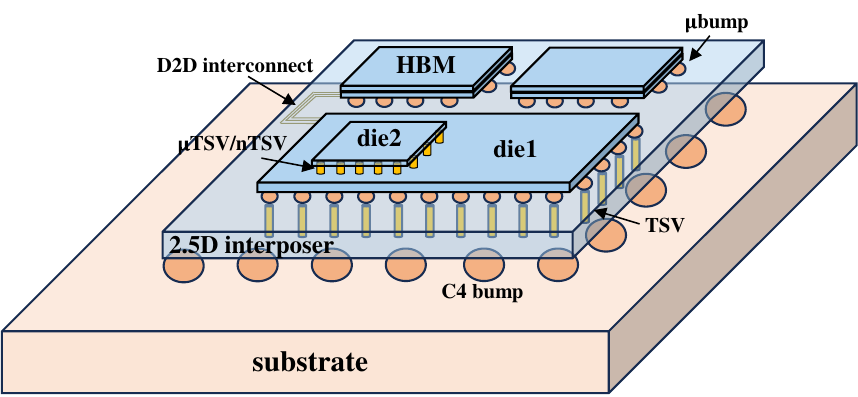}
    \caption{Schematic of a typical 2.5D/3D IC with multi-scale character. Numerous local fine structures are embedded in the system, such as TSVs, micro bumps, die-to-die interconnects, etc. These fine structures render full-system thermal stress simulation by conventional FEM extremely expensive.}
    \vspace{-15pt}
    \label{intro}
\end{figure}

However, 2.5D/3D ICs typically contain a significant number of local fine structures, such as through-silicon vias (TSVs), micro bumps, die-to-die interconnects, etc., as shown in Fig. \ref{intro}. These numerous fine structures embedded in the large systems lead to the multi-scale nature of 2.5D/3D ICs and renders full-system thermal stress analysis by conventional FEM extremely time- and memory-consuming, which is one of the key numerical challenges \cite{moroz20243dic}. Over the past decade, the linear superposition method has been widely used to perform fast estimations of thermal stress in 2.5D/3D ICs \cite{jung2012chip,li2013accurate,jung2014tsv,zhou2018thermal}. Although this method greatly reduces the analysis time, it is not a convergent numerical method and suffers from low accuracy, especially when the adjacent fine structures are very close or the local variations in the background stress are very sharp. Advanced numerical algorithms that are efficient, accurate, and flexible are urgently needed to tackle the multi-scalability challenge in full-system thermal stress analysis for 2.5D/3D ICs.

Fortunately, the most common fine structures in 2.5D/3D IC scenarios are generally configured in regular arrays for processing simplicity and reliability control, such as TSV arrays, micro bump arrays, C4 bump arrays, etc., where identical structures are repeated periodically \cite{lau2022recent,lau2023recent,pathak2011electromigration,jung2012chip}. The periodicity allows us to perform model order reduction to significantly reduce the total number of DoFs in the system. Meanwhile, the integration with sub-modeling, a technique which is commonly applied to refining a small part of the solution, can offer our algorithm with high flexibility in face of complex background stress.

Based on these novel ideas, we propose a strict numerical algorithm, named \textit{MORE-Stress}, aiming at reducing the cost of thermal stress simulation of large-scale arrays of fine structures in 2.5D/3D ICs and overcoming the multi-scalability numerical challenge. In this paper, we apply our algorithm to the thermal stress simulation of TSV arrays, one of the most common and important fine structures in 2.5D/3D ICs, to demonstrate our algorithm's core principles and representative performances. It is worth noting that our algorithm is not limited by TSVs and is adaptable to other types of fine structures in 2.5D/3D ICs, such as micro bumps, pillars, direct bondings, etc., regardless of their geometries. The key contributions of this work are as follows:
\begin{itemize}
    \item We propose a novel strict numerical algorithm based on model order reduction, which remarkably improves the computational efficiency for thermal stress simulation of TSV arrays in 2.5D/3D ICs. The algorithm consists of a one-shot local stage, where reduced order models of the TSV structures are developed, and a global stage, where the thermal stress of TSV arrays with arbitrary array sizes and thermal loads can be efficiently calculated.
    \item We design a procedure for calculating the thermal stress of TSV arrays embedded at arbitrary locations in a package system in combination with the sub-modeling technique, which makes our algorithm highly flexible in various scenarios.
    \item Extensive experimental results show that our algorithm can realize a 153--504$\times$ reduction in computational time and a 39--115$\times$ reduction in memory usage compared with the commercial FEM software ANSYS, with negligible errors less than 1$\%$. Our algorithm is as efficient as the linear superposition method, with an order of magnitude smaller errors and fast convergence.
\end{itemize}

The rest of this paper is organized as follows. Section 2 reviews the relevant work. Section 3 formulates the studied problem. Section 4 demonstrates the framework and principles of our algorithm. Section 5 provides the experimental setup and shows the results. Section 6 concludes this paper. The code has been released on GitHub.

%\vspace{-.05in}
\titlespacing*{\section}{0pt}{2ex plus .0ex minus .0ex}{.3ex plus .0ex}
\vspace{-5pt}
\section{Related Work}
\label{sec:review}
For the past decade, the linear superposition method has been widely utilized to perform fast thermal stress estimations for 2.5D/3D ICs, especially for the TSV arrays \cite{jung2012chip,jung2014tsv}. The idea of this method is to perform high-fidelity FEM simulation to get the stress tensors for each single structure and then superpose the stress tensors directly. Although this method greatly reduces the analysis time, it overlooks the coupling between adjacent structures and the local variations in the global background stress, which leads to a lack of accuracy especially when the pitches are small or when the background stress changes sharply.

Several works are trying to improve the accuracy of the linear superposition method or find other solutions. Li et al. \cite{li2013accurate} proposed a two-stage semi-analytical method that considers the interactions between nearby TSVs to provide more accurate analysis results. However, this method cannot be generalized to more complicated structures for it is analytical, and there still exists certain errors because the authors only considered interactions between pairs of TSVs. Zhou et al. \cite{zhou2018thermal} proposed a novel adaptive strategy finite element method to better simulate the stress distribution of a single TSV and then utilize the naive linear superposition method to get the stress distribution of the whole TSV array. This work is still within the framework of the linear superposition method and does not make fundamental improvements. Besides, Wang et al. \cite{wang2019runtime} have recently proposed a deep-learning-based method for the run-time thermal stress analysis of TSV arrays. However, the reliability of the deep-learning methods cannot be guaranteed and their model is limited to the training scenarios, so we do not consider similar deep-learning-based methods in this work.

%\vspace{-.05in}
\titlespacing*{\section}{0pt}{2ex plus .0ex minus .0ex}{.3ex plus .0ex}
\vspace{-1.0ex}
\section{Problem Formulation}
\label{sec:formulation}
\subsection{Governing Equation of Thermal Stress}
The governing equation of thermal stress is as follows \cite{hetnarski2009thermal}:
\begin{equation} \label{eq1}
\vspace{-5pt}
\begin{split}
-\nabla\cdot\sigma{(\boldsymbol{u})}&=\boldsymbol{f}\quad \rm{in}\quad\Omega\,, \\
\sigma{(\boldsymbol{u})} &= \lambda\rm{tr}(\epsilon(\boldsymbol{u}))\cdot\rm{\boldsymbol{1}}+2\mu\epsilon(\boldsymbol{u})-\alpha(3\lambda+2\mu)\cdot\Delta{\mathit{T}}\cdot\rm{\boldsymbol{1}}\,, \\
\epsilon(\boldsymbol{u}) &= \frac{1}{2}(\nabla{\boldsymbol{u}}+(\nabla{\boldsymbol{u}})^{T})\,,
\end{split}
\end{equation}
where $\Omega$ is the computational domain, $\boldsymbol{u}$ is the three-dimensional displacement vector field, $\epsilon(\boldsymbol{u})$ is the strain tensor field, $\sigma(\boldsymbol{u})$ is the stress tensor field, $\boldsymbol{f}$ is the body force vector, $\lambda$, $\mu$ are the Lam\'{e} parameters, $\alpha$ is the thermal expansion coefficient, $\Delta{\mathit{T}}$ is the thermal load between room temperature and package processing temperature, and $\rm{\boldsymbol{1}}$ represents the second-order unit tensor. The displacement vector field $\boldsymbol{u}$, the strain tensor field $\epsilon$, and the stress tensor field $\sigma$ are all functions of the space coordinate $\boldsymbol{r}=(x, y, z)^T$. In short, the first equation is the equilibrium equation, the second equation is the constitutive law, and the third equation is the strain-displacement relation.

The Lam\'{e} parameters are related with the common Young modulus $E$ and Poisson's ratio $\nu$ as follows:
\vspace{-5pt}
\begin{equation} \label{eq2}
\vspace{-5pt}
        \lambda = E\nu/(1+\nu)/(1-2\nu),\quad\mu = E/2/(1+\nu)\,.
\end{equation}

\subsection{FEM for Thermal Stress Simulation}
The simulation of thermal stress belongs to the computational structural mechanics, which is generally and most suitably performed by FEM. Following the convention, our algorithm is based on FEM as well, so we briefly introduce the FEM formulation for thermal stress simulation in this section \cite{larson2013finite}.

The first step of FEM is to convert the governing equation into the weak form, or the integral form. In the scenarios of integrated circuits, the gravity is ignored so the body force $\boldsymbol{f}$ is zero everywhere, and the boundary surfaces are generally assumed to be free or clamped so there are only Dirichlet boundary conditions. Under these assumptions, the weak form of Eq. \ref{eq1} is:
\vspace{-5pt}
\begin{equation} \label{eq3}
    a(\boldsymbol{u},\boldsymbol{v}) = L(\boldsymbol{v})\,,
\end{equation}
where $\boldsymbol{u}$ is called the trial function, and $\boldsymbol{v}$ is called the test function. The bilinear form $a(u,v)$ is equal to:
\begin{equation} \label{eq4}
\vspace{-5pt}
    a(\boldsymbol{u},\boldsymbol{v}) = \int_{\Omega}(\lambda\rm{tr}(\epsilon(\boldsymbol{u}))\cdot\rm{\boldsymbol{1}}+2\mu\epsilon(\boldsymbol{u})):\epsilon(\boldsymbol{v})\rm{d}\boldsymbol{r}\,,
\end{equation}
where the symbol $:$ denotes the contraction of tensors. The linear form $L(\boldsymbol{v})$ is equal to:
\begin{equation} \label{eq5}
\vspace{-5pt}
    L(\boldsymbol{v}) = \int_{\Omega}(\alpha(3\lambda+2\mu)\cdot\Delta{\mathit{T}}\cdot\rm{\boldsymbol{1}}):\epsilon(\boldsymbol{v})\rm{d}\boldsymbol{r}\,.
\end{equation}

After obtaining the weak form, the computational domain is discretized into a mesh, and $\boldsymbol{u}$, $\boldsymbol{v}$ are substituted by linear combinations of shape functions $\boldsymbol{\varphi}_i$ associated with each node $i$ of the mesh with the coefficient $\alpha_i$. By doing so, the weak form (Eq. \ref{eq3}) is assembled into a system of linear equations:
\begin{equation} \label{eq6}
    A\alpha = b\,,
\end{equation}
where $A$ is called the stiffness matrix and $b$ is called the load vector. After solving Eq. \ref{eq6}, the final solution of the thermal stress problem expressed in displacement is:
\begin{equation} \label{eq7}
    \boldsymbol{u} = \sum_{i}\boldsymbol{\varphi}_i\alpha_i\,,
\end{equation}
and then the strain $\epsilon$ and the stress $\sigma$ can be calculated based on Eq. \ref{eq1}. 

If the system is large and the size of the mesh is small, the number of DOFs (the dimension of $\alpha$) will be very large and the assembling and solving of Eq. \ref{eq6} will be extremely time- and memory-consuming, which is the case for thermal stress simulation of TSV arrays in 2.5D/3D ICs. Advanced numerical algorithms have to be designed to decrease the number of DoFs to be solved and reduce the computational cost.

\titlespacing*{\section}{0pt}{1.2ex plus .0ex minus .0ex}{.3ex plus .0ex}
\begin{figure}[t]
    \centering
    \includegraphics[width=0.20\textwidth]{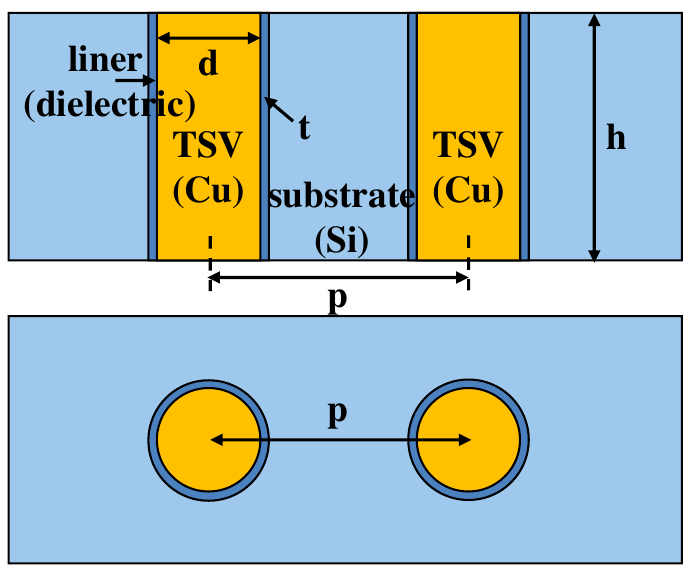}
    \caption{Sectional view and top view of the TSV structure adopted in this paper \cite{li2013accurate}. It consists of a copper TSV body in the silicon substrate and a dielectric liner. $d$ denotes the diameter of the TSV, $h$ denotes the height, $t$ denotes the thickness of the liner, and $p$ denotes the pitch of adjacent TSVs.}
    \vspace{-1.5em}
    \label{schematic}
\end{figure}

\begin{figure*}[htbp]
    \centering
    \includegraphics[width=0.80\textwidth]{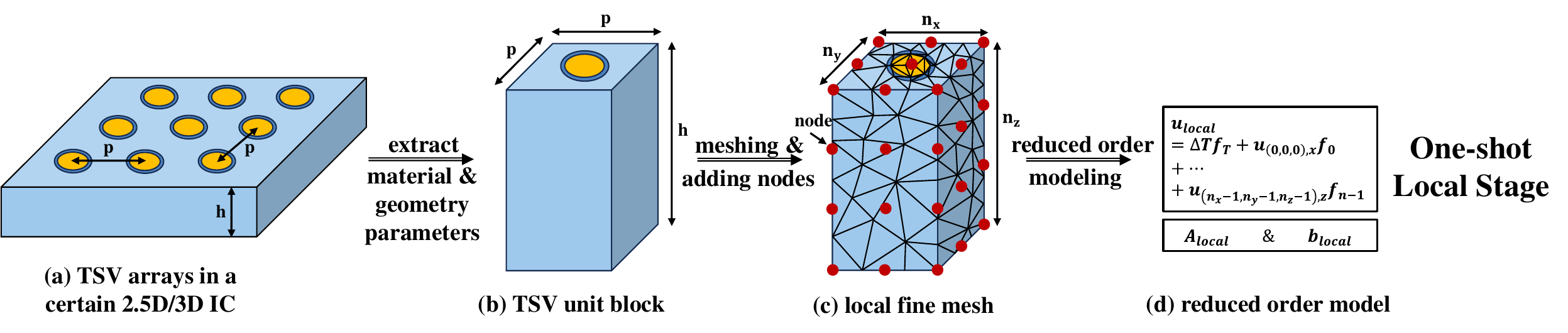}
    \caption{Illustration of the one-shot local stage, which will be performed only once for a certain set of material and geometry parameters. (a)(b) The material and geometry parameters of the TSV structures to be studied in a certain 2.5D/3D IC are extracted and the unit TSV block is set up. (c) Lagrange interpolation points are added to the surface of the unit block for reduced order modeling, and a fine mesh is developed for the unit block to assemble the local problem. (d) Reduced order model of the unit TSV block is obtained and ready to be applied to the global stage.}
    \label{algo_local}
    \vspace{-10pt}
\end{figure*}
\begin{figure*}[htbp]
    \centering
    \includegraphics[width=0.80\textwidth]{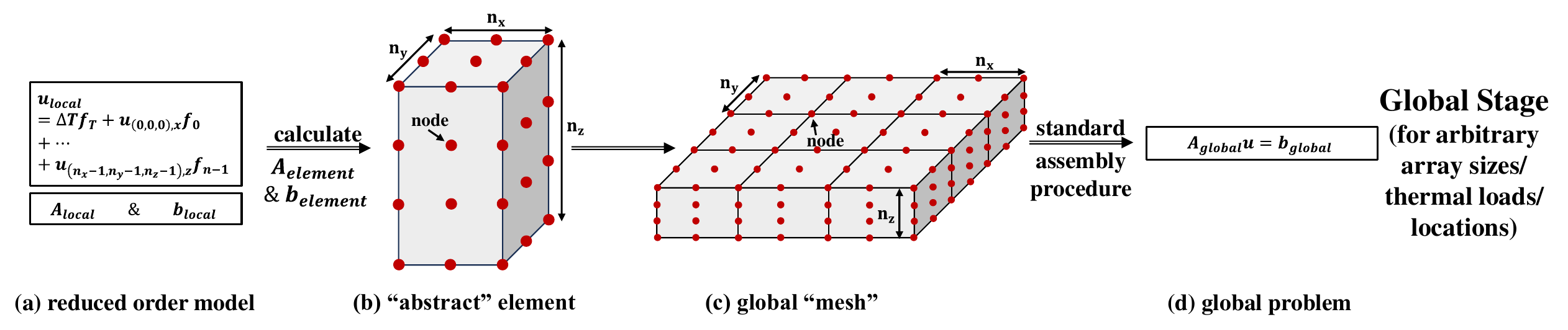}
    \caption{Illustration of the global stage, which will performed every time given a new problem. Once the one-shot local stage is performed, thermal stress of TSV arrays with arbitrary array sizes, under arbitrary thermal loads, and at arbitrary locations in a package system can be efficiently calculated in the global stage. (a) The corresponding pre-calculated reduced order model is loaded. (b)--(d) A standard assembly procedure is applied to assemble the global problem.}
    \label{algo_global}
    \vspace{-10pt}
\end{figure*}
\section{The MORE-Stress Algorithm}
In this section, we elaborate on the framework and principles of the proposed \textit{MORE-Stress} algorithm, through the case of TSV arrays.
\vspace{-5pt}
\subsection{Overall Framework}
Our algorithm consists of a \textbf{one-shot local stage} and a \textbf{global stage}. For a certain set of material and geometry parameters, the local stage only needs to be performed once, which typically finishes within minutes. After finishing the one-shot local stage, the thermal stress of TSV arrays with arbitrary array sizes, under arbitrary thermal loads, and at arbitrary locations in a package system can be calculated with huge speedup and memory usage reduction during the global stage.

In the \textbf{one-shot local stage} (see Sec .\ref{sec:local}), the TSV unit block is constructed based on the material and geometry parameters of the TSV structures to be studied, and a series of local problems associated with the unit block are solved to get the local basis functions for the global problem. In the \textbf{global stage} (see Sec. \ref{sec:global}), the global stiffness matrix and global load vector are assembled using the local basis functions, the local stiffness matrix, and the local load vector calculated in the one-shot local stage. After assembling, the generated global problem is solved and then the displacement field and stress field are calculated by a linear combination of the local basis functions.

In our algorithm, the reduction of computational cost is rooted in the reduction of the number of DoFs. The number of DoFs of the fine mesh for a unit block is reduced to the number of local basis functions. Because the TSV arrays are periodic and the TSV unit blocks are identical, the reduced order modeling can be performed just once in the local stage and applied to every unit block in the global stage to generate a model with much fewer DoFs, which is the philosophy of our algorithm. Obviously, the methodology can be generalized to any repeated, array-like fine structures in 2.5D/3D ICs, and the complexity of the geometry of the structure only affects the computational time of the one-shot local stage.

In the next subsections, we will detailedly explain the principles of the one-shot local stage and the global stage of our algorithm, and its combination with the sub-modeling technique. 
\vspace{-2.0ex}
\subsection{The One-shot Local Stage}
\label{sec:local}
As discussed above, the complexity of the geometry of the local structure does not affect the performance of our algorithm, so for brevity, we choose a simplified TSV structure which consists of the copper TSV body and a dielectric liner as shown in Fig. \ref{schematic} \cite{li2013accurate}. The geometric parameters include the height $h$ and the diameter $d$ of the copper body, the thickness $t$ of the liner, and the pitch $p$ of adjacent TSVs. 

A TSV array can be regarded as consisting of identical TSV unit blocks due to its periodicity, as shown by Fig. \ref{algo_local}(a)(b). A TSV unit block consists of a single TSV structure in the middle of a silicon substrate cuboid, whose dimension is $p\times p \times h$. In this paper, we only study the 90$^{\circ}$ TSV arrays, while the unit blocks of 120$^{\circ}$ TSV arrays can be chosen as a hexagonal prism with a TSV structure in the middle. 

After defining the unit block, a group of equally spaced nodes is placed on the corners and surfaces of the unit block as shown in Fig. \ref{algo_local}(c), serving as the Lagrange interpolation points for the displacement field of the unit block boundaries. The boundary displacement is approximated by the Lagrange interpolation functions defined on the nodes, which is the only source of error in our algorithm. Its convergence is guaranteed by the convergence of Lagrange interpolation.

The numbers of nodes along the three axes are denoted by $(n_x, n_y, n_z)$. For the node $(i, j, k)$ at coordinate $(x_i, y_j, z_k)$, the Lagrange interpolation function $L_{3D}(\boldsymbol{r}; i, j, k)$ is written as:
\begin{equation} \label{eq8}
    L_{3D}(\boldsymbol{r}; i, j, k) = L_{1D}(x; i)\times L_{1D}(y; j) \times L_{1D}(z; k)\,,
\end{equation}
where $L_{1D}(x; i)$ is the 1D Lagrange interpolation function on node $i$ along the x axis:
\begin{equation} \label{eq9}
    L_{1D}(x; i) = \frac{(x-x_0)\cdots(x-x_{i-1})(x-x_{i+1})\cdots(x-x_{n_x-1})}{(x_i-x_0)\cdots(x_i-x_{i-1})(x_i-x_{i+1})\cdots(x_i-x_{n_x-1})}\,,
\end{equation}
and the same is for $L_{1D}(y; j)$ and $L_{1D}(z; k)$. The boundary displacement of the unit block can then be approximated by the Lagrange interpolation functions $L_{3D}$ and the displacement of the nodes $\boldsymbol{u}_{i,j,k}$:
\begin{equation} \label{eq10}
\begin{split}
    &u_{bc, x}(\boldsymbol{r}) \approx \sum_{(i, j, k)}u_{(i, j, k), x}L_{3D}(\boldsymbol{r}; i, j, k)\,, \\
    &u_{bc, y}(\boldsymbol{r}) \approx \sum_{(i, j, k)}u_{(i, j, k), y}L_{3D}(\boldsymbol{r}; i, j, k)\,, \\ 
    &u_{bc, z}(\boldsymbol{r}) \approx \sum_{(i, j, k)}u_{(i, j, k), z}L_{3D}(\boldsymbol{r}; i, j, k)\,, \\
\end{split}
\end{equation}
where the x, y, and z components are listed separately for clarity.

After discretizing the unit block with a fine mesh as shown in Fig. \ref{algo_local}(c), the Dirichlet boundary problem (where boundary displacement is assigned) of the unit block (Eq. \ref{eq3}) is assembled into:
\begin{equation} \label{eq11}
    A_{local}\alpha_{local} = \Delta T b_{local}\,,
\end{equation}
as Eq. \ref{eq6}, or in a detailed form:
\begin{equation} \label{eq12}
    \begin{pmatrix} A_{local,f,f} & A_{local,f,bc} \\ A_{local,bc,f} & A_{local,bc,bc} \end{pmatrix}\begin{pmatrix}
        \alpha_{local,f} \\ \alpha_{local,bc}
    \end{pmatrix} = \Delta T \begin{pmatrix} b_{local,f} \\ b_{local,bc} \end{pmatrix}\,,
\end{equation}
where the subscript $f$ denotes the free DoFs and the subscript $bc$ denotes the boundary DoFs. To assign the boundary displacement to the boundary DoFs, a procedure called ``lifting'' is performed. The rows of the stiffness matrix corresponding to the boundary DoFs are set to zeros except that the diagonal elements are set to ones, and the elements of the load vector corresponding to the boundary DoFs are set to the preassigned displacement. Finally, we get the lifted system of linear equations that determine the free DoFs:
\begin{equation} \label{eq13}
    A_{local,f,f}\alpha_f = \Delta T b_{local,f}-A_{local,f,bc}u_{local,bc}\,.
\end{equation}
Substituting Eq. \ref{eq10} into Eq. \ref{eq13}, we can get the approximated, or the order-reduced local problem:
\begin{equation} \label{eq14}
\vspace{-5pt}
\begin{split}
    &A_{local,f,f}\alpha_{local,f} \approx \Delta T b_{local,}f - A_{local,f,bc}L\times \\
    &\begin{pmatrix}
        u_{(0,0,0),x} & u_{(0,0,0),y} & u_{(0,0,0),z} & \cdots & u_{(n_x-1,n_y-1,n_z-1),z}
    \end{pmatrix}^T \,,
\end{split}
\end{equation}
where $L$ is the matrix containing the geometric information of Lagrange interpolation functions, which is solved automatically and does not need to be explicitly calculated. 

Using the Lagrange interpolation, we successfully reduce the order of Eq. \ref{eq11} from the number of boundary DoFs to the number of interpolation nodes. The displacement field of the unit block can then be written as:
\begin{equation} \label{eq15}
\vspace{-5pt}
\begin{split}
    &\boldsymbol{u}_{local} \approx \Delta T \boldsymbol{f}_T + u_{(0,0,0),x}\boldsymbol{f}_0 + u_{(0,0,0),y}\boldsymbol{f}_1 + u_{(0,0,0),z}\boldsymbol{f}_2 + \cdots + \\
    &u_{(n_x-1,n_y-1,n_z-1),z}\boldsymbol{f}_{n-1}\,,
\end{split}
\end{equation}
where $\boldsymbol{f}_{i}$ is the solution of the local problem setting the thermal load $\Delta T$ to zero, the corresponding components of nodal displacement to one and all other components to zeros, while $\boldsymbol{f}_T$ is the solution setting the thermal load $\Delta T$ to one, and all the components of nodal displacement to zeros. The total number of $\boldsymbol{f}_i$s, $n$, is equal to:
\begin{equation} \label{eq16}
\vspace{-5pt}
    n = \{n_x\times n_y \times n_z - (n_x-2)\times (n_y-2) \times (n_z-2)\}\times 3\,.
\end{equation}
All of the $\boldsymbol{f}_i$s, together with the $\boldsymbol{f}_T$, constitute the set of local basis functions as shown in Fig. .\ref{algo_local}(d). The total number of the local basis functions is much smaller than the number of DoFs in the fine mesh of the unit TSV block, which is the root of model order reduction. 

Because the stiffness matrix is the same for all of the local problems ($A_{local,f,f}$), the time-consuming LU or Cholesky decomposition needs to be performed only once and the intermediate results can be reused for all of the local problems, which greatly reduces the computational time of the one-shot local stage. Meanwhile, the calculation of local problems can be easily parallelized on the task level, which further reduces the time cost.

\subsection{The Global Stage}
\label{sec:global}
Once the one-shot local stage is finished, the thermal stress of TSV arrays (with the same material and geometric parameters) with arbitrary array sizes and under arbitrary thermal loads can be quickly calculated in the global stage. 

In the global stage, the unit TSV block can be regarded as an abstract ``element'' with its own nodal displacement components $u_{(0,0,0),x},\cdots$ $,u_{(n_x-1, n_y-1, n_z-1),z}$ being the element DoFs, while the TSV array to be simulated can be regarded as 
an abstract ``mesh'' made up by abstract ``elements'' with the entire nodal displacement components being the global DoFs, as shown in Fig. \ref{algo_global}(b)(c). In this perspective, the global problem has no difference from a common finite element problem. The global stiffness matrix $A_{global}$ and the global load vector $b_{global}$ can be easily assembled through the standard assembly procedure \cite{larson2013finite}.

The remaining work is to calculate the element stiffness matrix $A_{element}$ and element load vector $b_{element}$ which are requested by the standard assembly procedure. Notice that $\boldsymbol{f}_i$, the local basis function corresponding to the $i$th element DoF, can be expressed as the linear combination of the local shape functions $\boldsymbol{\varphi}_{local}$ as Eq. \ref{eq7}:
\begin{equation} \label{eq17}
\vspace{-5pt}\boldsymbol{f}_i=\sum_p\boldsymbol{\varphi}_{local,p}\alpha_p\,.
\end{equation}
Considering Eq. \ref{eq3} and \ref{eq4}, the $(i,j)$th entry of $A_{element}$ is calculated by:
\begin{equation} \label{eq18}
\vspace{-5pt}
\begin{split}
    &A_{element,i,j}=a(\boldsymbol{f}_i, \boldsymbol{f}_j) \\
    &=\sum_p\sum_q \alpha_p\times a(\boldsymbol{\varphi}_{local,p},\boldsymbol{\varphi}_{local,q})\times \alpha_q = \alpha^T A_{local}\alpha\,,
\end{split}
\end{equation}
where $\alpha=\begin{pmatrix} \alpha_0 & \alpha_1 & \cdots & \alpha_n\end{pmatrix}^T$. Considering Eq. \ref{eq3} and \ref{eq5}, the $i$th entry of $b_{element}$ is calculated by:
\begin{equation} \label{eq19}
\vspace{-5pt}
    \begin{split}
        &b_{element,i}=L(\boldsymbol{f}_i) \\
        &=\sum_p \alpha_p\times L(\boldsymbol{\varphi}_{local,i}) = \alpha^T\Delta T b_{local}\,.
    \end{split}
\end{equation}
$A_{element}$ is an $n\times n$ dense array, and $b_{element}$ is an $n$-dimensional vector, where $n$ is the number of element DoFs calculated by Eq. \ref{eq16}.

After getting the element stiffness matrix $A_{element}$ and the element load vector $b_{element}$, a standard finite element assembly procedure can be performed to get the global counterparts $A_{global}$ and $b_{global}$. The detailed process is omitted here for brevity. $A_{global}$ is a sparse matrix because only nodes in the same elements (unit blocks) or shared by adjacent elements (unit blocks) can contribute to its entries.
 
With $A_{global}$ and $b_{global}$ prepared, the global problem is to be solved:
\begin{equation} \label{eq20}
\vspace{-5pt}
    A_{global}u=b_{global}\,,
\end{equation}
as shown in Fig. \ref{algo_global}(d), where $u$ is the vector consisting of the nodal displacement components. Eq. \ref{eq20} is better solved by iterative methods such as GMRES for a shorter computational time because we do not need to solve the same equation repeatedly in the global stage.

Finally, the displacement field $\boldsymbol{u}(\boldsymbol{r})$ of a certain unit block can be calculated using the corresponding entries of $u$ and the local basis functions $\boldsymbol{f}_0,\boldsymbol{f}_1,\cdots,\boldsymbol{f}_{n-1}$ together with $\boldsymbol{f}_{T}$ following Eq. \ref{eq15}. Further the strain $\epsilon(\boldsymbol{r})$ and stress $\sigma(\boldsymbol{r})$ can be calculated by derivation of the displacement routinely.
\vspace{-1.0ex}
\subsection{Combination with Sub-modeling}
\label{sec:sub}
Sub-modeling is a common technique in the engineering practice of structural mechanics \cite{wong2022warpage} and is widely applied in commercial software like ANSYS . If a small part of a large system is of interest, a coarse mesh is first developed for the whole system and a coarse solution is obtained. Then the part of interest is cut out to form a sub-model. After that, a fine mesh is solely developed for the sub-model, and the coarse solution is applied to its boundaries as boundary conditions. This procedure avoids discretizing the whole system with the fine mesh and is proved to be accurate if the boundary of the sub-model is far away enough from the part of interest in engineering.

Our algorithm is naturally compatible with sub-modeling. The displacement values obtained from the coarse solutions can be directly assigned to the boundary nodes of the TSV arrays through the ``lifting'' procedure as described by Eq. \ref{eq11}, \ref{eq12} and \ref{eq13}. Moreover, to satisfy the condition that the boundaries of the sub-model should be far away enough from the part of interest, arbitrary ``dummy'' unit blocks can be added to the periphery of the TSV array. A ``dummy'' unit block has the same dimension and nodal distribution as a TSV unit block, but it is a pure silicon cuboid without a TSV structure in the middle. An extra local stage has to be performed for the ``dummy'' unit block, and the corresponding $A_{element,dummy}$ and $b_{element,dummy}$ are calculated in the same way as Eq. \ref{eq18} and \ref{eq19}. The standard assembly procedure can handle hybrid elements without difficulty \cite{larson2013finite}.

The compatibility with sub-modeling implies that our algorithm can be easily integrated with other stress simulators, commercial or open-sourced, and can be treated as a module specializing in the local fine structures in 2.5D/3D ICs in a comprehensive simulation flow.
%\vspace{-.05in}
\titlespacing*{\section}{0pt}{1.5ex plus .0ex minus .0ex}{.3ex plus .0ex}
\section{Experimental Setup and Results}
\label{sec:experiment}
\begin{figure}[t]
    \centering
    \subfloat[]{
        \includegraphics[width=0.3\textwidth]{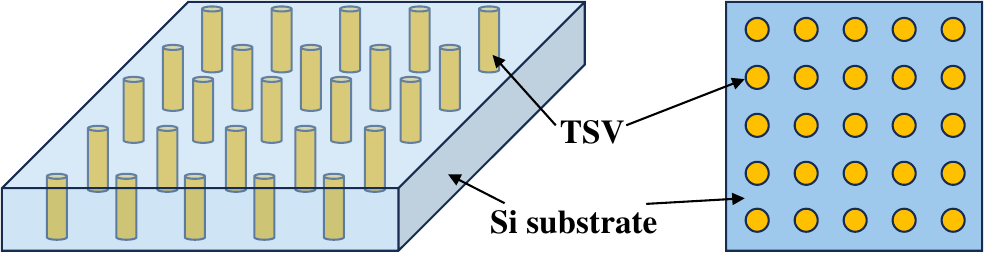}
        \label{intro_a}
    } \vskip -2pt
    \subfloat[]{
        \includegraphics[width=0.3\textwidth]{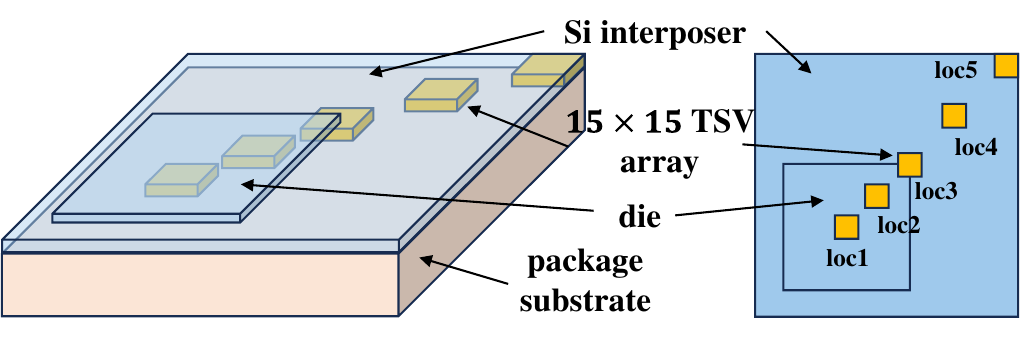}
        \label{intro_b}
    } \vskip -1pt
    \caption{(a) The first scenario. TSV arrays with array sizes ranging from $10\times10$ to $50\times50$ are studied. Except for efficiency and accuracy, this scenario is designed to test the scalability and convergence of our algorithm. (b) The second scenario is a $15\times15$ TSV array embedded at five different locations in a chiplet. This scenario is designed to test the combination of our algorithm with the sub-modeling technique.}
    \label{scenario}
    \vspace{-15pt}
\end{figure}

\subsection{Implementation}
Our algorithm is implemented based on an open-sourced FEM framework, FeniCSx \cite{barrata2023dolfinx,BarattaEtal2023,ScroggsEtal2022,BasixJoss,AlnaesEtal2014}, in Python. Besides, we use PETSc \cite{balay2019petsc} as the linear algebra backend and use Gmsh \cite{geuzaine2009gmsh} to mesh the unit TSV block in the local stage. The algorithm is designed to support parallelization, both for the one-shot local stage and the global stage. The experiments are conducted on a Linux server with an Intel Xeon Silver 4210R 2.40GHz processor (20 logical CPUs). To balance the threading overhead and the gain of parallelization, we set the number of threads to 16 for the one-shot local stage and the number of threads to 8 for the global stage hereafter.
\vspace{-5pt}
\subsection{Experimental Setup}
Two representative scenarios are studied to test the performance of our algorithm, as shown in Fig. \ref{scenario}. The first scenario is a group of standalone TSV arrays with array sizes ranging from $10\times10$ to $50\times50$ and with their top and bottom surfaces clamped \cite{jung2014tsv,zhou2018thermal}. This scenario is designed to test the convergence and scalability of our algorithm. The height $h$ and diameter $d$ of the copper via is set to be 50 $\upmu$m and 5 $\upmu$m, respectively, and the thickness of the liner $t$ is set to be 500 nm, as shown in Fig. \ref{schematic} \cite{jung2014tsv,moroz20243dic}. Two pitches $p$ are tested in this scenario, which are 15 $\upmu$m and 10 $\upmu$m, respectively \cite{jung2014tsv}. The thermal load $\Delta T$ is set to be $-250$ $^{\circ}\rm{C}$ (annealing/reflow 275 $^{\circ}\rm{C}$ $\rightarrow$ room temperature 25 $^{\circ}\rm{C}$) to represent a fabrication process \cite{jung2012chip,lau2018warpage}.

The second scenario is a $15\times15$ TSV array embedded at five different locations in a chiplet, as shown in Fig .\ref{scenario}(b), which is designed to test the integration of our algorithm with the sub-modeling technique \cite{jung2012chip,moroz20243dic}. The stress of the TSV array couples with the global warpage stress of the chiplet in this scenario \cite{tsai2019theoretical}. The geometric parameters and the thermal load are the same as those in the first scenario. The pitch is set to be 15 $\upmu$m or 10 $\upmu$m, as well. The chiplet model consists of a composite substrate, a silicon interposer, and a silicon die. The TSV array is in the interposer. We first develop and solve a coarse model of the chiplet in ANSYS and then extract the sub-model and the displacement on the boundaries of the sub-model from the coarse solution. Two columns and two rows of ``dummy'' unit blocks are added to the edge of the TSV array to keep enough distance between the boundaries and the TSV array, as explained in Sec. \ref{sec:sub}.

Following the convention, we calculate the gridded von Mises stress on the cut plane crossing the half height of the TSV arrays for comparison \cite{jung2012chip,jung2014tsv,zhou2018thermal}. The number of grid points in a single TSV unit block is set as $100\times 100$ to capture full details of stress distribution. For each case, the \textbf{mean absolute error} (MAE) between the result provided by our algorithm and the ground truth is calculated and normalized by the maximum von Mises stress because the stress is in direct proportion to the thermal load. We record the runtime of the global stage as the computational time of our algorithm, because the local stage only needs to be performed once for the same material and geometry parameters, and its typical runtime is within minutes, which is very short for large-scale analysis. For memory usage, we record the maximum memory usage during computation.

All simulations are performed in the commercial software ANSYS on a cloud server with an Intel Xeon Platinum 8350C 2.60GHz processor (24 logical CPUs). Because the models studied have large numbers of DoFs, we set the solver type to ``iterative'' in ANSYS during simulations.
\begin{table}[t]
\renewcommand{\arraystretch}{1.1}
\centering
\setlength{\tabcolsep}{0.8mm}{
\begin{tabular}{|c|c|c|c|c|c|c|}
\hline
\multicolumn{7}{|c|}{$p=15$ $\upmu$m} \\
\hline
\hline
\multicolumn{2}{|c|}{array size} & $10\times10$ & $20\times20$ & $30\times30$ & $40\times40$ & $50\times50$ \\
\hline
\multirow{2}{*}{ANSYS} & time &391 s& 1296 s& 2952 s& 5392 s& 8612 s\\
\cline{2-7}
 & memory & 12.2 G& 53.1 G& 119.7 G& 212.7 G& 330.1 G\\
\hline
Linear & time &2.4 s & 3.9 s& 9.6 s& 13.4 s& 20.5 s\\
\cline{2-7}
superposition & memory & 0.24 G& 0.47 G& 0.81 G& 1.23 G& 1.72 G\\
 \cline{2-7}
\cite{jung2012chip,jung2014tsv} & error & 5.36$\%$ & 6.57$\%$ & 7.27$\%$ & 7.69$\%$ & 7.98$\%$\\
\hline
\multirow{3}{*}{Ours} & time & 2.5 s& 4.4 s& 9.5 s& 13.1 s& 17.1 s\\
\cline{2-7}
 & memory & 0.31 G & 0.67 G & 1.23 G & 1.85 G & 2.91 G\\
 \cline{2-7}
 & error & 0.93$\%$ & 0.87$\%$ & 0.80 $\%$ & 0.72 $\%$ & 0.67 $\%$ \\
\hline
Improve. & time & 156$\times$& 295$\times$ & 311$\times$ & 412$\times$& 504$\times$\\
\cline{2-7}
over ANSYS & memory & 39$\times$ & 79$\times$ & 97$\times$ & 115$\times$ & 113$\times$\\
\hline 
Improve. & & & & & & \\
over Linear& accuracy & 5.8$\times$ & 7.6$\times$ & 9.1$\times$ & 10.7$\times$ & 12.0$\times$\\
superposition & & & & & &\\
\hline
\hline
\multicolumn{7}{|c|}{$p=10$ $\upmu$m} \\
\hline
\hline
\multicolumn{2}{|c|}{array size} & $10\times10$ & $20\times20$ & $30\times30$ & $40\times40$ & $50\times50$ \\
\hline
\multirow{2}{*}{ANSYS} & time & 352 s& 1201 s& 2840 s& 5304 s& 8508 s\\
\cline{2-7}
 & memory & 12.6 G & 48.0 G& 108.2 G & 196.8 G & 312.9 G\\
\hline
Linear & time & 2.3 s& 4.0 s& 9.6 s& 13.3 s& 20.7 s\\
\cline{2-7}
superposition & memory &0.24 G & 0.46 G& 0.81 G& 1.23 G& 1.71 G\\
 \cline{2-7}
\cite{jung2012chip,jung2014tsv} & error & 7.40$\%$ & 10.33$\%$ & 12.37$\%$ & 13.60$\%$ & 14.43$\%$\\
\hline
\multirow{3}{*}{Ours} & time & 2.3 s& 4.5 s& 11.4 s& 13.9 s& 18.5 s\\
\cline{2-7}
 & memory & 0.31 G& 0.67 G& 1.22 G& 1.86 G& 2.90 G\\
 \cline{2-7}
 & error & 0.98 $\%$ & 0.92 $\%$ & 0.85 $\%$ & 0.79 $\%$ & 0.74 $\%$ \\
\hline
Improve. & time & 153$\times$ &  267$\times$ & 250$\times$ & 382$\times$ & 460$\times$\\
\cline{2-7}
over ANSYS & memory & 41$\times$ & 72$\times$ & 87$\times$ & 106$\times$ & 108$\times$\\
\hline
Improve. & & & & & & \\
over Linear & accuracy &7.6$\times$ & 11.2$\times$ & 14.6$\times$ & 17.2$\times$ & 19.5$\times$ \\
superposition & & & & & & \\
\hline
\end{tabular}}
\caption{Summarized computational time, memory usage, and computational errors of ANSYS, linear superposition, and our algorithm for the first scenario (Fig. \ref{scenario}(a)). The computational errors do not apply to ANSYS because its results are the ground truth. The improvement in computational time and memory usage of our algorithm over ANSYS and the improvement in accuracy of our algorithm over the linear superposition method are also listed.}
\label{scene1}
\vspace{-20pt}
\end{table}
\vspace{-20pt}
\subsection{Experimental Results and Discussion}
\subsubsection{Efficiency $\&$ Accuracy.}
The efficiency and accuracy of our algorithm in the two scenarios are detailedly studied in this section. We set the number of interpolation nodes on the surface of the unit block to be $(4,4,4)$ to balance efficiency and accuracy in the following experiments. For the first scenario, the one-shot local stage costs 301.6 seconds and 287.4 seconds for the $p=15$ $\upmu$m case and $p=10$ $\upmu$m case, respectively. The computational time and memory usage of ANSYS, the linear superposition method, our algorithm, and the computational errors (normalized MAEs) are summarized in Table \ref{scene1}. Our algorithm realizes a great reduction in computation time and memory usage compared with ANSYS. The speedup of our algorithm ranges from 153--504$\times$, and the reduction in memory usage ranges from 39--113$\times$. The improvement increases with the size of the TSV array because there are some fixed costs like the I/O time, which implies that our algorithm can realize more than 500$\times$ reduction in simulation time and more than 100$\times$ reduction in memory usage if the code is carefully optimized. The errors of our algorithm are less than 1$\%$, which is negligible considering the huge improvement in efficiency. Compared with the linear superposition method, our algorithm is as efficient (and even slightly more efficient when the task scale is large), with an order of magnitude smaller errors. The linear superposition method works badly for the $p=10$ $\upmu$m case because it overlooks the coupling between adjacent TSVs, while the accuracy of our algorithm is basically not affected. Moreover, the computational errors of our algorithm decrease with the size of the TSV array because the errors concentrate near the boundaries. On the contrary, the errors of the linear superposition method increase with the size of the TSV array because its errors are distributed over the whole computational domain. As the part of interest is the TSV array itself, this is another manifestation that our algorithm prevails over the linear superposition method in accuracy.
\begin{table}[h]
\vspace{-10pt}
\renewcommand{\arraystretch}{1.1}
    \centering
    \caption{Summarized results for the second scenario (Fig .\ref{scenario}(b)).}
    \setlength{\tabcolsep}{1.3mm}{
    \begin{tabular}{|c|c|c|c|c|c|c|}
\hline
\multicolumn{7}{|c|}{$p=15$ $\upmu$m} \\
\hline
\hline
\multicolumn{2}{|c|}{location} & loc1 & loc2 & loc3 & loc4 & loc5 \\
\hline
\multirow{2}{*}{ANSYS} & time & 1137 s & 1078 s& 1006 s& 1020 s& 1012 s\\
\cline{2-7}
 & memory & 37.6 G & 38.2 G & 37.4 G& 37.6 G& 38.0 G\\
\hline
Linear & time & 3.4 s& 3.3 s& 3.4 s& 3.5 s& 3.5 s\\
\cline{2-7}
superposition & memory & 0.39 G& 0.40 G& 0.40 G& 0.39 G& 0.40 G\\
 \cline{2-7}
\cite{jung2012chip,jung2014tsv} & error & 5.62$\%$& 5.60$\%$& 7.23$\%$& 5.70$\%$& 7.40$\%$\\
\hline
\multirow{3}{*}{Ours} & time & 3.5 s& 3.5 s& 3.6 s& 3.5 s& 3.7 s\\
\cline{2-7}
 & memory & 0.46 G& 0.47 G& 0.46 G& 0.46 G& 0.46 G\\
 \cline{2-7}
 & error & 0.62$\%$ & 0.62$\%$ & 0.71$\%$ & 0.71$\%$ & 0.72$\%$\\
\hline
Improve. & time & 325$\times$ & 308$\times$ & 279$\times$ & 291$\times$ & 273$\times$\\
\cline{2-7}
over ANSYS & memory & 82$\times$ & 81 $\times$ & 81$\times$ & 82$\times$ & 83$\times$\\
\hline
Improve. & & & & & & \\
over Linear & accuracy & 9.1$\times$ & 9.0$\times$ & 10.2$\times$ & 8.0$\times$ & 10.3$\times$\\
superposition & & & & & & \\
\hline
\hline
\multicolumn{7}{|c|}{$p=10$ $\upmu$m} \\
\hline
\hline
\multicolumn{2}{|c|}{location} & loc1 & loc2 & loc3 & loc4 & loc5 \\
\hline
\multirow{2}{*}{ANSYS} & time & 1086 s & 1042 s& 964 s& 980 s& 963 s\\
\cline{2-7}
 & memory & 36.4 G & 37.2 G & 37.0 G& 36.8 G& 37.1 G\\
\hline
Linear & time & 3.3 s& 3.4 s& 3.4 s& 3.4 s& 3.5 s\\
\cline{2-7}
superposition & memory & 0.39 G& 0.40 G& 0.39 G& 0.40 G& 0.39 G\\
 \cline{2-7}
\cite{jung2012chip,jung2014tsv} & error & 6.62$\%$& 6.60$\%$& 8.23$\%$& 6.70$\%$& 8.79$\%$\\
\hline
\multirow{3}{*}{Ours} & time & 3.6 s& 3.5 s& 3.7 s& 3.6 s& 3.7 s\\
\cline{2-7}
 & memory & 0.45 G& 0.46 G& 0.46 G& 0.45 G& 0.46 G\\
 \cline{2-7}
 & error & 0.67$\%$ & 0.68$\%$ & 0.75$\%$ & 0.73$\%$ & 0.78$\%$\\
\hline
Improve. & time & 302$\times$& 298$\times$& 261$\times$& 272$\times$& 260$\times$\\
\cline{2-7}
over ANSYS & memory & 81$\times$ & 81$\times$ & 80$\times$ & 82$\times$ & 81$\times$\\
\hline
Improve. & & & & & & \\
over Linear & accuracy & 9.9$\times$ & 9.7$\times$ & 11.0$\times$ & 9.2$\times$ &11.3$\times$ \\
superposition & & & & & & \\
\hline
    \end{tabular}}
    \label{scene2}
    \vspace{-10pt}
\end{table}

For the second scenario, the computational time, memory usage, and computational errors (normalized MAEs) of ANSYS, our algorithm, and linear superposition method are summarized in Table \ref{scene2}. The one-shot local stage has been performed when studying the first scenario, and it does not need to be performed again because the geometry does not change. When the TSV array is close to or at the locations where the background stress changes sharply, such as the corner of the chip (loc3) and the corner of the interposer (loc5) as shown in Fig \ref{scenario}(b), the errors of the linear superposition method are large. On the contrary, the accuracy of our algorithm is almost unaffected by this phenomenon because it follows the standard sub-modeling procedure, and the local variations in the global background stress are captured by the displacement boundary conditions assigned to the boundary nodes.
\begin{table}[h]
\vspace{-20pt}
\renewcommand{\arraystretch}{1.1}
\centering
\caption{Convergence of our algorithm. $n$ refers to the number of element DoFs (Eq. \ref{eq16}). The simulation time of ANSYS for this case is 1296 s.}
\setlength{\tabcolsep}{1.3mm}{
\begin{tabular}{|c|c|c|c|c|c|}
\hline
 $(n_x, n_y, n_z)$ & $(2,2,2)$ & $(3,3,3)$ & $(4,4,4)$ & $(5,5,5)$ & $(6,6,6)$ \\
\hline
\hline
$n$ & 24 & 78 & 168 & 294 & 456 \\
\hline
one-shot& \multirow{2}{*}{147.0 s} & \multirow{2}{*}{204.1 s} & \multirow{2}{*}{301.6 s} & \multirow{2}{*}{431.8 s} & \multirow{2}{*}{603.2 s}\\
local stage runtime & & & & &\\
\hline
global stage runtime & 2.2 s & 2.6 s & 4.4 s& 12.8 s& 25.1 s\\
\hline
error & 5.25$\%$ & 2.07$\%$& 0.87$\%$& 0.44$\%$& 0.28$\%$\\
\hline
\end{tabular}}
\label{convergency}
\vspace{-10pt}
\end{table}

\begin{figure}[h]
    \centering
    \includegraphics[width=0.45\textwidth]{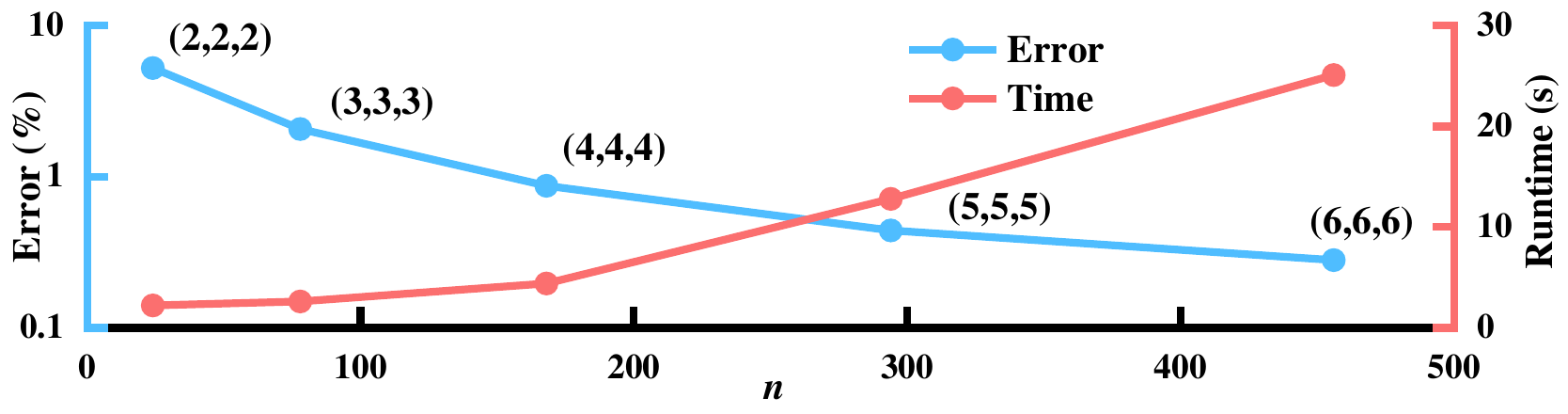}
    \vspace{-2pt}
    \caption{Computational errors and runtime (of the global stage) plotted against numbers of element DoFs, $n$. $(n_x,n_y,n_z)$s are labelled beside the corresponding points. The vertical axis of error is in log scale.}
    \label{convergence}
    \vspace{-15pt}
\end{figure}
\subsubsection{Convergence.}
Next, we study the convergence of our algorithm on a $20\times20$ standalone TSV array with $p=15$ $\upmu$m. The simulation time of ANSYS for this case is 1296 s. We set the numbers of nodes used in the one-shot local stage from $(2, 2, 2)$ to $(6, 6, 6)$. The corresponding numbers of element DoFs $n$, computational errors (normalized MAEs), and computational time are summarized in Table \ref{convergency}. Computational errors and runtime (of the global stage) are also plotted against $n$, in Fig. \ref{convergence}. It is clear that as the number of element DoFs $n$ increases, the computational errors decrease rapidly, implying that our algorithm enjoys fast convergence. This is because as the number of Lagrange interpolation nodes increases, the interpolation functions can better fit the surface displacement of the unit block. On the contrary, there is no concept of convergence for the linear superposition method because its accuracy is not influenced by any algorithm parameters.

%\vspace{-.05in}
\section{Conclusion}
\label{sec:Conclusion}
In this paper, we propose a novel strict numerical algorithm \textit{MORE-Stress}, aiming at accelerating thermal stress simulation of large-scale arrays of fine structures in 2.5D/3D ICs and overcoming the multi-scalability numerical challenge. Our algorithm is based on FEM, and utilizes the periodicity of local fine structures to realize a great reduction in the number of DoFs through model order reduction. Meanwhile, combination with the sub-modeling technique makes our algorithm highly flexible. In this work, we focus on and apply our algorithm to the thermal stress simulation of TSV arrays in various scenarios. Extensive experimental results demonstrate that our algorithm can realize a 153--504$\times$ reduction in simulation time and a 39--115$\times$ reduction in memory usage compared with the commercial FEM software ANSYS, with negligible errors less than 1$\%$. Our algorithm is as efficient as the linear superposition method, with an order of magnitude smaller errors and fast convergence. The proposed algorithm is general and adaptable to different types of fine structures in 2.5D/3D ICs, such as mircro bumps, pillars, direct bondings, etc., regardless of their geometries. In the future, we plan to apply it to more types of local fine structures and extend it to more complicated scenarios.

\clearpage

\vspace{-.05in}
{
%\scriptsize
\small
\bibliographystyle{IEEEtran}
\bibliography{./ref/Top_sim,./ref/Stress}
}

\end{document}